\documentclass[usenatbib]{mn2e}

\usepackage{graphicx}
\usepackage{amsmath}
\usepackage{lscape}
\usepackage{afterpage}
\usepackage{color, xcolor}
\usepackage{soul}

\title[Irradiated accretion discs in ULXs]{Irradiated, colour-temperature-corrected accretion discs in ultraluminous X-ray sources}

\author[A. D. Sutton, C. Done and T. P. Roberts]{Andrew D. Sutton\thanks{Email: andrew.sutton@durham.ac.uk}, Chris Done, Timothy P. Roberts\\
 \\
 Department of Physics, University of Durham, South Road, Durham, DH1 3LE, UK
}

\pagerange{\pageref{firstpage}--\pageref{lastpage}} \pubyear{2014}

\def\Msun{\hbox{$M_{\odot}$}}

\def\H0{{\rm ~km~s^{-1}~Mpc^{-1}}}

\def\la{\mathrel{\hbox{\rlap{\hbox{\lower4pt\hbox{$\sim$}}}{\raise2pt\hbox{$<$}}
}}}
\def\ga{\mathrel{\hbox{\rlap{\hbox{\lower4pt\hbox{$\sim$}}}{\raise2pt\hbox{$>$}}
}}}
\def\d25{$D_{25}$}

\begin{document}

\maketitle

\label{firstpage}

\begin{abstract}
Although attempts have been made to constrain the stellar types of optical counterparts to ULXs, the detection of optical variability instead suggests that they may be dominated by reprocessed emission from X-rays which irradiate the outer accretion disc.  Here, we report results from a combined X-ray and optical spectral study of a sample of ULXs, which were selected for having broadened disc-like X-ray spectra, and known optical counterparts.  We simultaneously fit optical and X-ray data from ULXs with a new spectral model of emission from an irradiated, colour-temperature-corrected accretion disc around a black hole, with a central Comptonising corona.  We find that the ULXs require reprocessing fractions of $\sim 10^{-3}$, which is similar to sub-Eddington thermal dominant state BHBs, but less than has been reported for ULXs with soft ultraluminous X-ray spectra.  We suggest that the reprocessing fraction may be due to the opposing effects of self-shielding in a geometrically thick super-critical accretion disc, and reflection from far above the central black hole by optically thin material ejected in a natal super-Eddington wind.  Then, the higher reprocessing fractions reported for ULXs with wind-dominated X-ray spectra may be due to enhanced scattering onto the outer disc via the stronger wind in these objects.  Alternatively, the accretion discs in these ULXs may not be particularly geometrically thick, rather they may be similar in this regard to the thermal dominant state BHBs.
\end{abstract}


\section{Introduction}

Ultraluminous X-ray sources (ULXs) are non-nuclear X-ray point sources in external galaxies, with X-ray luminosities in excess of $10^{39}~{\rm erg~s^{-1}}$.  They are brighter than the Eddington luminosity of a typical $\sim 10~\Msun$ Galactic black hole, so it was suggested that they may be powered by accretion on to a new population of intermediate mass black holes (IMBHs; $10^2$--$10^4~\Msun$; \citealt{colbert_and_mushotzky_1999}), with masses between those seen in stellar-remnant and supermassive regimes.  However, studies have demonstrated that the X-ray spectral and timing properties of a number of ULXs are unlike those seen in any of the known sub-Eddington accretion states, and are instead suggestive of a new accretion regime that occurs at mass accretion rates around, and in excess of the Eddington limit (e.g., \citealt{feng_and_soria_2011}, and references therein).  

In this case, ULXs may be powered by either typical stellar mass black holes (sMBHs; $<20~\Msun$; \citealt{feng_and_soria_2011}) or massive stellar black holes (MsBHs; 20--100 $\Msun$; \citealt{feng_and_soria_2011}), which may form from stellar collapse in regions of low metallicity (\citealt{zampieri_and_roberts_2009}; \citealt{belczynski_etal_2010}; \citealt{mapelli_etal_2010}).  Some ULX transients provide evidence in favour of stellar remnant black holes.  These include a microquasar in M31 \citep{middleton_etal_2014}, that displays highly variable radio emission when $L_{\rm X} \sim 10^{39}~{\rm erg~s^{-1}}$.  By comparison with Galactic black hole binaries (BHBs), this indicates that it is at around the Eddington luminosity.  This is further confirmed by the detection of a classic sub-Eddington thermal dominant (high/soft) state when the X-ray luminosity drops below $\sim 10^{39}~{\rm erg~s^{-1}}$.  Then, using the constraints on the Eddington luminosity, \cite{middleton_etal_2014} estimate a black hole mass of $\sim 10 \Msun$.  Also, \cite{liu_etal_2013} confirm a probable Wolf-Rayet companion star to a supersoft ULX in M101.  They use this to calculate a mass function, and find that it likely contains a $\sim$ 20--30 $\Msun$ black hole.  

\cite{gladstone_etal_2009} identified three types of ULX X-ray spectra, which they suggested may represent a spectral sequence with increasing accretion rate.  They suggested that around the Eddington limit ULXs appear with broad disc-like X-ray spectra, whilst at higher accretion rates a two component X-ray spectrum emerges.  In the two component spectra, a radiatively driven wind likely provides the soft component (\citealt{poutanen_etal_2007}; \citealt{kajava_and_poutanen_2009}), and the hard spectral component possibly originates from either a central cool, optically thick Comptonising corona, or the hot disc itself with a large colour correction (\citeauthor{middleton_etal_2011a} 2011a; \citealt{kajava_etal_2012}).  Although it was originally suggested that the precise balance of the two components depended on the accretion rate, with softer wind-dominated spectra occurring in the most super-Eddington sources, it now seems likely that the inclination of the system also plays a key role in determining the observed spectrum (e.g. \citealt{sutton_etal_2013b}).

Irrespective of the precise nature of the central black hole and its accretion state, ULXs must have a large reservoir of mass to accrete from, in order to produce their observed X-ray luminosities.  Indeed, point-like optical counterparts have been identified for a number of ULXs (\citealt{goad_etal_2002}; \citealt{liu_etal_2002}; \citealt{kaaret_etal_2004}; \citealt{liu_etal_2004}; \citealt{soria_etal_2005}; \citealt{kuntz_etal_2005}; \citealt{mucciarelli_etal_2005}; \citealt{liu_etal_2007}; \citealt{grise_etal_2008}; \citealt{roberts_etal_2008}; \citealt{rosado_etal_2008}; \citealt{grise_etal_2011}; \citealt{gladstone_etal_2013}), many of which are blue in colour, and appear to be consistent with large early-type companions; although a minority of the potential counterparts in the sample of \cite{gladstone_etal_2013} have colours which rule out O-type stars.  However, it may not be possible to trivially identify stellar counterparts to ULXs.  \cite{rappaport_etal_2005} caution that emission from the outer regions of the accretion disc may also contribute to the observed optical flux.  Furthermore, the optical light from the donor star in a ULX may be considerably less than would usually be expected for a non-donor star of the same initial mass, especially if mass transfer began early in its life. 

In Galactic BHBs the X-ray flux from the inner source irradiates, and is reprocessed in the outer regions of the accretion disc.  This can dominate the spectrum at optical and UV wavelengths (\citealt{van_paradijs_and_mcclintock_1995}; \citealt{gierlinski_etal_2009}).  The detection of stochastic variability superimposed onto the optical light curves of ULXs, most notably NGC 1313 X-2 (\citealt{grise_etal_2008}; \citealt{liu_etal_2009}; \citealt{impiombato_etal_2011}), indicates that X-ray reprocessing is occurring in ULXs.  This is possibly due to irradiation of the accretion disc or the companion star (e.g., \citealt{copperwheat_etal_2005}; \citealt{copperwheat_etal_2007}; \citealt{patruno_and_zampieri_2008}).   In general, both the disc and star contribute, with the relative contribution from the disc scaling as the cosine of the inclination.  But, in systems with large discs, the disc itself is more likely to dominate \citep{copperwheat_etal_2007}.   
Indeed, it has recently been argued that ULX counterparts are dominated by emission from the accretion disc (both intrinsic and reprocessed).  Evidence for this comes from a combination of the optical colours, variability and X-ray to optical flux ratios in the ULX sample of \cite{tao_etal_2011}, and the variable broad optical lines in NGC 1313 X-2 and Ho IX X-1, where the breadth of the lines indicate that the accretion discs must be close to face on \citep{roberts_etal_2011}.  Searches for associated X-ray and optical variability could provide a further test for reprocessed emission, as the irradiating flux would be dependent on the X-ray luminosity.  No current studies do this, perhaps because the required multiple epochs of simultaneous optical and X-ray observations would be difficult to schedule, and would require a large amount of total observing time.  

Spectral models of emission from an irradiated disc ({\sc diskir}; \citealt{gierlinski_etal_2008}; \citealt{gierlinski_etal_2009}) have been successfully used to model the non-contemporaneous broadband X-ray and optical spectra of a number of ULXs.  These fits indicate that a few per cent of the bolometric luminosity irradiates, and is reprocessed in, the outer accretion disc; e.g. 0.03 and 0.015 of the bolometric luminosities are reprocessed in the outer discs in NGC 5408 X-1 and NGC 6946 X-1 respectively (\citealt{grise_etal_2012}; \citealt{berghea_and_dudik_2012})\footnote{An irradiated fraction of $\sim 2 \times 10^{-3}$ is reported in ESO 243-49 HLX-1 \citep{soria_etal_2011}, although this source is arguably a poor comparator to the bulk of the ULX population, as it remains a good IMBH candidate.}.  These values are much higher than the irradiating fractions of $\sim 10^{-3}$ seen in thermal dominant state Galactic BHBs \citep{gierlinski_etal_2009}.  \cite{kaaret_and_corbel_2009} argue that such high reprocessing fractions rule out super-Eddington solutions in ULXs: geometrically thick accretion discs would result in self shielding in the disc; and in models invoking winds, they argue that the outer regions of the ULX do not see the central hard emission, rather they are only irradiated by soft photons from the outflow, which would not thermalise in the disc.  However, if a fraction of hard inner disc/coronal emission can be scattered by optically thin material in the wind much further from the ULX, then it may irradiate and thermalise in the outer regions of the accretion disc.

One limitation of the {\sc diskir} model is that it does not include a colour-temperature-correction \citep{done_etal_2012}.  In this work we use a new spectral model of emission from an irradiated, colour-temperature-corrected disc, with a central corona, to model the X-ray and optical data from several ULXs which were classified as broadened discs in \cite{sutton_etal_2013b}.  As it seems likely that optical counterparts to ULXs are dominated by the accretion disc, we here neglect any contribution from the donor star.

\section{A new spectral model of an irradiated accretion disc with a colour-temperature-correction}\label{optxirr}

Here, we use a new model of emission from black hole accretion to fit the X-ray and optical data from a ULX sample.  The new model - {\sc optxirr}\footnote{The authors expect to make the {\sc optxirr} model publicly available through {\sc xspec}.} - combines the colour-temperature-corrected accretion disc and energetically coupled corona of {\sc optxagnf} \citep{done_etal_2012} with an irradiated disc, as in {\sc diskir} \citep{gierlinski_etal_2008}.  {\sc diskir} models illumination of the outer disc by emission from the central disc and its Compton tail.  A fraction of the bolometric luminosity from the disc is thermalised to the local black body temperature at each radius, and contributes strongly to the UV and optical flux.  However, {\sc diskir} uses a multi-colour-disc prescription for the illuminating spectrum, and it neglects any colour-temperature-correction in the intrinsic disc emission.  {\sc optxagnf} \citep{done_etal_2012} is intended to model AGN spectra.  It approximates a colour-temperature-corrected accretion disc spectrum, and considers the system energetics self-consistently by including both a soft excess and high energy tail that are in energetic balance.  This is calculated by assuming that all radiation within a given radius does not thermalise, but is instead distributed between a power-law and low temperature Comptonised component.

{\sc optxirr} modifies the {\sc optxagnf} model to include irradiation of the outer disc by assuming that the disc is illuminated by a flux of $L_{\rm bol} / 4 \pi r^2$, a fraction of which ($f_{\rm out}$) is reprocessed by the disc.  The value of $f_{\rm out}$ depends on the geometry of the system, the albedo of the disc and the fraction of the non-reflected emission that is thermalised in the disc \citep{gierlinski_etal_2008}. In addition to $f_{\rm out}$, {\sc optxirr} has all of the parameters of {\sc optxagnf}: black hole mass, distance from the observer, Eddington ratio, black hole spin, the radius of the corona, the outer disc radius, the electron temperature and optical depth of the soft Comptonisation component, the spectral index of the power-law, the fraction of the emission below the coronal radius that is in the power-law, the redshift and normalisation.  

\begin{figure*}
\begin{center}
\includegraphics[width=15cm]{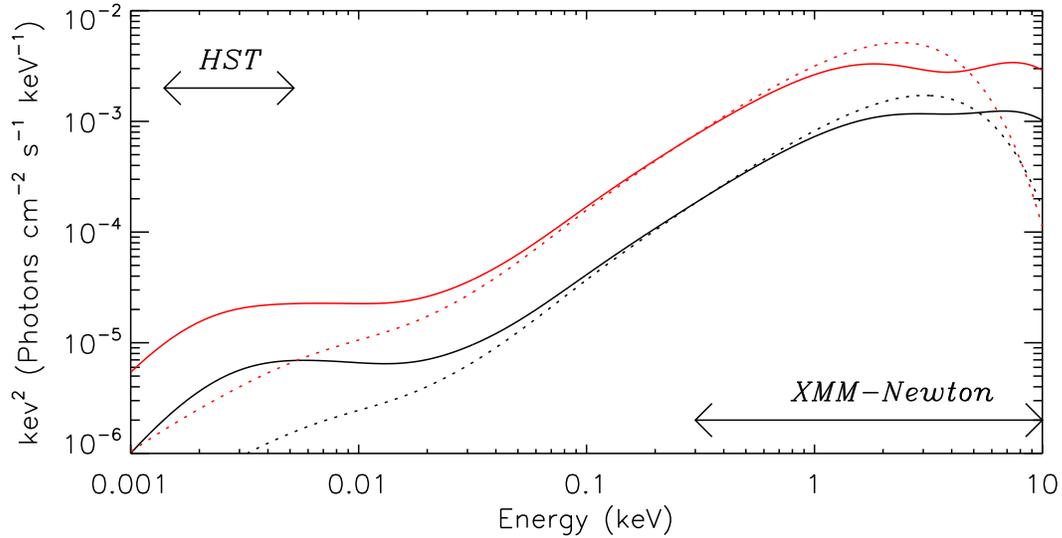}
\caption{Examples of model broad-band unabsorbed ULX spectra from the {\sc optxirr} spectral model, for a 10 (black) and 30 (red) $\Msun$ black hole at 3 Mpc distance from the observer, with ${\rm log} (L_{\rm bol}/L_{\rm Edd}) = 0.3$ and ${\rm log} (r_{\rm out}) = 6$.  The solid lines show spectra for ULXs with an optically-thick corona ($r_{\rm cor} = 25~r_{\rm g}$, $\tau = 20$, $kT_{\rm e}  = 2~{\rm keV}$), and a reprocessing fraction of $f_{\rm out} = 10^{-3}$; the dashed lines show the intrinsic colour-temperature-corrected disc spectra, with no corona or reprocessing in the outer disc; and, the two double sided arrows show the approximate energy coverage of the {\it HST} and {\it XMM-Newton} data analysed in this work.  The spectral curvature at $\sim$ 0.02 keV in the unirradiated disc spectra is due to the colour-temperature-correction.  Above the Lyman break hydrogen and helium are ionized, therefore electron scattering dominates the opacity, and the colour-temperature-correction is higher \citep{done_etal_2012}.}
\label{sample_spec}
\end{center}
\end{figure*}

\begin{figure*}
\begin{center}
\includegraphics[width=15cm]{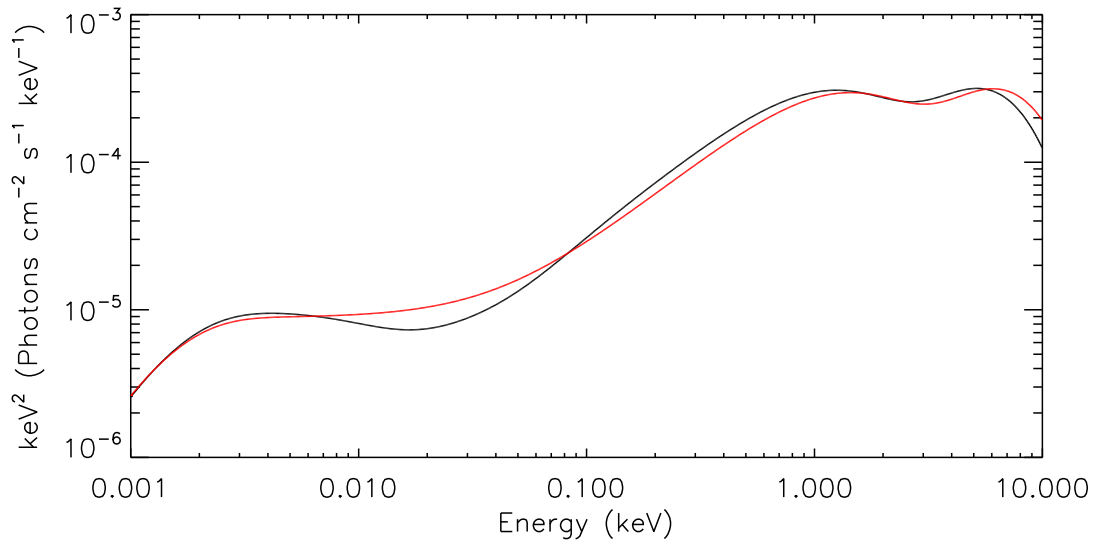}
\caption{A comparison of broadband spectral models -- {\sc optxirr} (black) and {\sc diskir} (red).  The optical and low flux X-ray data from M81 X-6 were fitted with the absorbed/reddened {\sc optxirr} model, as described in Section~\ref{data_analysis}, for a 30~{\Msun} black hole and an outer disc radius of $10^6~r_{\rm g}$.  This was repeated with the {\sc diskir} model, with the outer disc radius set equal to $1.67 \times 10^5~r_{\rm in}$, which corresponds to $10^6~r_{\rm g}$ for an inner disc radius of $6~r_{\rm g}$.  The absorption and reddening components were then removed from both models.  Critically, The reprocessing fractions resulting from the {\sc optxirr} and {\sc diskir} model fits are consistent with each other. These are $f_{\rm out} = 5^{+3}_{-2} \times 10^{-3}$ and $4 \pm 2 \times 10^{-3}$ respectively.}
\label{spec_compare}
\end{center}
\end{figure*}

A number of these parameters are not relevant to ULXs, and when fitting the model to ULX spectra we make a number of approximations: given the proximity of these sources ($d < 5~{\rm Mpc}$), we set the redshift to 0; we fix the distance to that of the ULX host galaxy; and, we set the flux in the hard power-law to zero, as such a component has not been detected in ULX spectra in the {\it XMM-Newton} EPIC band\footnote{Although such a component may be required to fit ULX spectra above $\sim$ 10 keV obtained with {\it NuSTAR} (\citealt{bachetti_etal_2013}; \citealt{walton_etal_2013}; \citealt{rana_etal_2014}; \citealt{walton_etal_2014}).}, instead all of the power below the coronal radius is in a cool Comptonised corona.  The model results in degeneracies in black hole spin, so we approximate the ULXs as having zero spin.  Additionally, we fix the model normalisation to 2 instead of 1, as is typically used for the {\sc optxagnf} model \citep{done_etal_2012}; this is appropriate for a face-on accretion disc, and differs by a factor $\sim 2$ from an inclination of 60 degrees.  Finally, it should be noted that {\sc optxagnf} (and {\sc optxirr}) neglects the drop in black hole efficiency from advection and mass loss in a wind, as the global mass accretion rate becomes super-critical.  Instead, it is assumed that the total luminosity is proportional to the mass accretion rate.  We show examples of spectra from the {\sc optxirr} model for a 10 and 30 $\Msun$ black hole in Figure~\ref{sample_spec}, and a comparison with the {\sc diskir} model in Figure~{\ref{spec_compare}}.  As well as the colour-temperature-correction, which mainly modifies the shape of the optical and UV spectrum, {\sc optxirr} differs from {\sc diskir} in several ways.  It includes a soft Comptonisation component, as well as the hard, power-law like Comptonisation component; it does not include inner disc irradiation by the Compton tail, which is important in the low/hard state \protect\citep{gierlinski_etal_2008}; and finally, the outer disc radius is defined in terms of $r_{\rm g}$, not the inner disc radius.

The physical scenario envisaged by {\sc optxirr} is of an accretion disc with a central corona.  This is broadly similar to the faintest sources in the original ultraluminous state description of ULX accretion, as suggested by \cite{gladstone_etal_2009}.  In this scenario, the two components in the X-ray spectrum correspond to emission from a soft truncated disc and a hard corona.  However, current thinking is that super-Eddington accretion results in massive radiatively driven outflows \citep{poutanen_etal_2007}, which are critical in producing the two component spectrum observed in the brightest ULXs.  
Furthermore, in sources with strong outflows, the inner-edge of the wind may itself be irradiated by X-ray emission from the central disc, and also contribute to the optical/UV flux \citep{vinokurov_etal_2013}.  
However, below $\sim 3 \times 10^{39}~{\rm erg~s^{-1}}$ the ULX population is dominated by sources with broad, disc-like X-ray spectra \citep{sutton_etal_2013b}.  Although observations demonstrate that a number of these have subtle two component X-ray spectra, the interpretation of the two components remains unclear.  Possibly these faintest ULXs represent a transition phase in the evolution of ULXs at around the Eddington accretion rate, where an optically thick corona  obscures the central disc (e.g. \citeauthor{middleton_etal_2011b} 2011b; \citealt{middleton_etal_2012}).  However, \cite{dotan_and_shaviv_2011} suggest that characteristic super-Eddington-like properties would emerge as the Eddington rate is approached, in which case even the faintest ULXs may have emerging `classic' ultraluminous spectra, with the soft X-ray emission originating in the photosphere of a wind.

Regardless of the precise nature of the X-ray emitting region in the inner disc, the outer disc is locally sub-Eddington.  Then, depending on the geometry of the system, this can be irradiated with some X-ray spectrum, which is reprocessed to the local black body temperature.  As such, the reprocessing fraction that we estimate with {\sc optxirr} does not depend strongly on the physics occuring in the central disc, as long as the model provides a reasonable approximation of the mass accretion rate, and bolometric flux.  
However, if there is strong mass loss in an outflow, the mass accretion rate estimated from the X-ray spectrum may be an underestimate of the global value.  As we then apply this mass accretion rate to the outer regions of the disc, we may underestimate the intrinsic emission from the outer disc.  Potentially, this could in turn lead us to attribute too much of the optical/UV emission to X-ray reprocessing.

\section{Sample selection \& data reduction}

\begin{table*}
\centering
\caption{ULX sample}
\begin{tabular}{lccc}
\hline
Source name & 2XMM/CXO ID & ${N_{\rm H}}^a$ & Distance$^b$ \\
\hline
NGC 253 ULX2  & 2XMM J004732.9-251749 & 1.38 & 3.68$^i~~$ \\
NGC 1313 X-2  & 2XMM J031822.1-663603 & 3.95 & 4.39$^{ii~}$ \\
NGC 2403 X-1  & 2XMM J073625.5+653540 & 4.17 & 3.50$^i~~$ \\
M81 X-6       & 2XMM J095532.9+690034 & 4.22 & 4.27$^i~~$ \\
NGC 4190 ULX1 & CXO  J121345.2+363754 & 1.62 & 3.47$^{iii}$ \\
\hline
\end{tabular}
\label{sample}
\begin{minipage}{\linewidth}
$^a$Galactic column density in the direction of the ULX ($\times 10^{20}~{\rm cm^{-2}}$), from \cite{dickey_and_lockman_1990}.
$^b$Distance to the ULX host galaxy in Mpc, taken from: $^i$\cite{tully_etal_2009}; $^{ii}$\cite{EDD2}; $^{iii}$\cite{tikhonov_and_karachentsev_1998}.
\end{minipage}
\end{table*}

\subsection{Sample selection}

The model that we use for this study - {\sc optxirr} - does not consider the effects of radiatively driven winds or advection on the emitted spectrum.  Such effects are thought to occur in super-Eddington sources (e.g. \citealt{poutanen_etal_2007}).  As such, we want to avoid using the model to fit strongly super-Eddington sources.  Instead, we consider the `broadened disc' ULXs \citep{sutton_etal_2013b}, which are suspected to have mass accretion rates close to the Eddington limit.  To carry out broadband spectral modelling, we require both X-ray and optical data.  Thus, we selected all of the ULXs with broadened disc spectra in \cite{sutton_etal_2013b}, that also had optical counterparts identified by \cite{gladstone_etal_2013} in {\it HST} Hubble Legacy Archive (HLA\footnote{\tt http://hla.stsci.edu/}) data.  This resulted in a sample of 5 ULXs, details of which are shown in Table~\ref{sample}.  For 4 of the 5 sources in the sample, \cite{gladstone_etal_2013} identified only a single potential optical counterpart in the HLA images; the exception to this was NGC 1313 X-2, which had 2 potential counterparts.  Here, we use only the counterpart to NGC 1313 X-2 previously identified by \cite{liu_etal_2007}, and confirmed by \cite{grise_etal_2008}.

\subsection{{\em XMM-Newton} data reduction}

\begin{table*}
\centering
\caption{{\it XMM-Newton} observation log}
\begin{tabular}{cccccc}
\hline
Obs. ID$^a$ & Date$^b$ & Exposure time$^c$ & $\theta ^d$ & ${f_{\rm X}}^e$ & flux group$^f$ \\
\hline
\multicolumn{6}{c}{{\bf NGC 253 ULX2}} \\
0125960101 & 2000-06-03 & 31.6/36.3/35.7 & 1.7 & 1.5 & - \\
0125960201 & 2000-06-04 & 5.6/9.1/9.0 & 1.6 & 2.1 & - \\
0110900101 & 2000-12-13 & 8.2/26.7/29.1 & 3.1 & 1.9 & - \\
0152020101 & 2003-06-19 & 61.9/79.5/80.4 & 1.6 & 1.5 & - \\
0304851101 & 2005-12-16 & 7.4/19.9/19.8 & 0.7 & 1.1 & - \\
0304850901 & 2006-01-02 & 8.8/11.4/11.4 & 0.8 & 1.2 & - \\
0304851001 & 2006-01-06 & 8.8/11.4/11.4 & 0.8 & 1.2 & - \\
0304851201 & 2006-01-09 & 15.3/19.4/19.4 & 0.8 & 1.2 & - \\
0304851301 & 2006-01-11 & 4.0/6.8/6.8/ & 0.8 & 1.2 & - \\
\\
\multicolumn{6}{c}{{\bf NGC 1313 X-2}} \\
0106860101 & 2000-10-17 & 18.8/27.1/26.9 & 5.4 & 0.8 & low \\
0150280101 & 2003-11-25 & 1.3/2.5/2.5 & 1.0 & 2.0 & high \\
0150280301 & 2003-12-21 & 7.5/10.1/10.2 & 1.0 & 3.4 & high \\
0150280401 & 2003-12-23 & 3.2/5.4/5.7 & 1.0 & 3.9 & high \\
0150280501 & 2003-12-25 & 2.3/8.5/8.7 & 1.0 & 1.7 & low \\
0150280601 & 2004-01-08 & 7.2/11.2/11.4 & 1.0 & 1.1 & low \\
0150281101 & 2004-01-16 & 3.3/7.2/7.3 & 1.1 & 1.0 & low \\
0205230201 & 2004-05-01 & -/7.7/8.0 & 1.3 & 0.8 & low \\
0205230301 & 2004-06-05 & 8.8/11.5/11.5 & 1.3 & 4.0 & high \\
0205230401 & 2004-08-23 & 4.4/10.6/10.7 & 1.1 & 0.8 & low \\
0205230501 & 2004-11-23 & 12.5/15./15.5 & 1.0 & 0.9 & low \\
0205230601 & 2005-02-07 & 8.7/11.9/11.9 & 1.1 & 3.9 & high \\
0301860101 & 2006-03-06 & 17.4/21.3/21.3 & 6.2 & 3.6 & high \\
0405090101 & 2006-10-15 & 80.0/105.3/105.4 & 5.4 & 3.4 & high \\
\\
\multicolumn{6}{c}{{\bf NGC 2403 X-1}} \\
0150651101 & 2003-04-30 & -/5.6/5.7 & 1.7 & 1.2 & - \\
0150651201 & 2003-09-11 & -/5.1/5.4 & 3.8 & 1.3 & - \\
0164560901 & 2004-09-12 & 51.2/70.6/72.6 & 5.4 & 1.2 & - \\
\\

\multicolumn{6}{c}{{\bf M81 X-6}} \\
0111800101 & 2001-04-22 & 74.5/-/96.3 & 3.4 & 2.8 & high \\
0112521001 & 2002-04-10 & 7.0/10.0/10.0 & 11.9 & 2.8 & high \\
0112521101 & 2002-04-16 & 7.6/9.9/10.0 & 11.9 & 2.8 & high \\
0200980101 & 2004-09-26 & -/99.4/101.0 & 13.9 & 4.0 & high \\
0657802001 & 2011-03-24 & 3.3/-/7.1 & 6.5 & 1.1 & low \\
0657801801 & 2011-09-26 & 6.0/14.9/14.0 & 7.8 & 3.0 & high \\
0657802201 & 2011-11-23 & 13.0/16.0/16.1 & 8.2 & 1.2 & low \\
\\
\multicolumn{6}{c}{{\bf NGC 4190 ULX1}} \\
0654650101 & 2010-06-06 & 0.9/7.7/7.5 & 1.1 & 2.9 & low \\
0654650201 & 2010-06-08 & 8.2/17.2/18.8 & 1.1 & 3.0 & low \\
0654650301 & 2010-11-25 & 9.3/14.8/14.8 & 1.2 & 5.6 & high \\
\hline
\end{tabular}
\begin{minipage}{\linewidth}
$^a${\it XMM-Newton} observation identification number.
$^b$Date of the {\it XMM-Newton} observation (yyyy-mm-dd).
$^c$The amount of good time for each EPIC detector (pn/MOS1/MOS2) in units of ks.
$^d$Off-axis angle of the source in the EPIC detectors in arcminutes.
$^e$Approximate observed 0.3-10 keV flux in units of $\times 10^{-12}~{\rm erg~cm^{-2}~s^{-1}}$, calculated from an absorbed {\it p}-free disc model in {\sc xspec}.  The luminosity is calculated from the pn data where available, or the MOS1 data in observations with no pn detection.
$^f$The X-ray flux bin in which the observation was included for subsequent spectral fitting.  Dashes are shown for sources in which flux binning was not used.
\end{minipage}
\label{xmm_obs_log}
\end{table*}

\begin{table*}
\centering
\caption{{\it HST} observation log}
\begin{tabular}{ccccc}
\hline
Date$^a$ & Instrument & Filter & Exposure time$^b$ & Magnitude$^c$ \\
\hline
\multicolumn{5}{c}{{\bf NGC 253 ULX2}} \\
2006-09-13 & ACS-WFC & F475W & 1482 & $21.7 \pm 0.3$ \\
2006-09-13 & ACS WFC & F606W & 1508 & $20.8 \pm 0.2$ \\
2006-09-13 & ACS-WFC & F814W & 1534 & $19.6 \pm 0.2$ \\
\\
\multicolumn{5}{c}{{\bf NGC 1313 X-2}} \\
2003-11-22 & ACS-HRC & F330W & 2760 & $21.6 \pm 0.8$ \\
2003-11-22 & ACS-WFC & F435W & 2520 & $23.2 \pm 0.4$ \\
2003-11-22 & ACS-WFC & F555W & 1160 & $23.4 \pm 0.5$ \\
2003-11-22 & ACS-WFC & F814W & 1160 & $23.5 \pm 0.6$ \\
\\
\multicolumn{5}{c}{{\bf NGC 2403 X-1}} \\
2005-10-17 & ACS-HRC & F330W & 2912 & $23 \pm 2$ \\
2005-10-17 & ACS-WFC & F435W & 1248 & $25 \pm 1$ \\
2005-10-17 & ACS-WFC & F606W & 1248 & $24.5 \pm 0.6 $ \\
\\
\multicolumn{5}{c}{{\bf M81 X-6}} \\
1995-01-31 & WFPC2   & F336W & 1160 & $22 \pm 2$ \\
2006-03-22 & ACS-WFC & F435W & 1200 & $23.7 \pm 0.6$ \\
2001-06-04 & WFPC2   & F555W & 8000 & $24 \pm 2$ \\
2006-03-22 & ACS-WFC & F606W & 1200 & $23.8 \pm 0.6$ \\
2001-06-04 & WFPC2   & F814W & 8000 & $23 \pm 2$ \\
\\
\multicolumn{5}{c}{{\bf NGC 4190 ULX1}} \\
2008-01-07 & WFPC2 & F300W & 4400 & $23.0 \pm 0.3$ \\
2008-01-07 & WFPC2 & F450W & 4400 & $24 \pm 2$ \\
2008-01-08 & WFPC2 & F606W & 1600 & $24 \pm 2$ \\
2008-03-21 & WFPC2 & F814W & 4400 & $25 \pm 3$ \\
\hline
\end{tabular}
\begin{minipage}{\linewidth}
$^a$Data of the {\it HST} observation (yyyy-mm-dd).  $^b$Exposure time of the {\it HST} observation in units of s.  $^c$Apparent Vega magnitude of the source corrected for Galactic reddening, taken from Gladstone et al. (2013).
\end{minipage}
\label{hst_obs_log}
\end{table*}

Having defined the sample, we then obtained all of the available {\it XMM-Newton} EPIC data from observations of the regions containing the 5 ULXs from the HEASARC data archive\footnote{\tt http://heasarc.gsfc.nasa.gov/}.  All of the {\it XMM-Newton} observations included in this study are summarised in Table~\ref{xmm_obs_log}.  The data were reduced, and standard products extracted using {\it XMM-Newton} {\sc sas}\footnote{\tt http://xmm.esa.int/sas/} (version 11.0.0).  Firstly, high energy (10-15 keV) full-field light curves were extracted using {\sc xmmselect} with {\tt FLAG} $=0$ and {\tt PATTERN} $\le 4$ or 12 for pn and MOS detections respectively.  Several observations were excluded from further analysis as they contained high levels of solar flaring throughout the high energy light curves, these were: 0150280201 (NGC 1313 X-2), 0150280701 (NGC 1313 X-2), 0657801601 (M81 X-6).  Of the remaining observations, the majority contained periods of high background flaring\footnote{Exceptions to this were: all EPIC detectors in  observations 0304850901, 0304851001, 0205230501 and 0112521001; the MOS1 and MOS2 detectors in observations 0304851201, 0205230301 and 0301860101; plus the pn detector in observation 0112521101.}, which were filtered out using good time interval files created using {\sc tabgtigen}.  The exact filtering count rate varied between observations, to maximise the utilised data whilst avoiding flaring, but typical values were $\sim 1.5$ and $\sim 0.6~{\rm ct~s^{-1}}$ for the pn and MOS detectors respectively.

For 4 of the 5 ULXs, source spectra were typically extracted from circular apertures with radii 30--50 arcseconds; however, for observations of NGC 253 ULX2 a neighbouring source necessitated a very small source region of 12.5 arcseconds.  Also, in a few cases elliptical source regions were used when the source neighboured a detector chip gap.  Background spectra were extracted from larger circular regions located at a similar distance from the read-out node as the source, on the same or a neighbouring chip in the same quadrant for pn detections, or on the same chip as the source for MOS detections.  Source and background spectra were created using {\sc ogip spectral products} in {\sc xmmselect}, with {\tt FLAG} $=0$ and {\tt PATTERN} $\le 4$ or 12 for pn and MOS data respectively.  Finally, the data were binned to a minimum of 20 counts per bin using the {\sc ftool}\footnote{\tt https://heasarc.gsfc.nasa.gov/ftools/} {\sc grppha}.

\subsection{{\em HST} magnitudes}

{\it HST} magnitudes for the sample ULXs were taken from \cite{gladstone_etal_2013}.  These were derived from an assortment of ACS-HRC, ACS-WFC and WFPC2 observations over a range of dates (see Table~\ref{hst_obs_log} for details). The Galactic-extinction-corrected Vega magnitudes (Table~4 of \citealt{gladstone_etal_2013}) were converted into units of $\rm photons~cm^{-2}~s^{-1}$.  When converting the WFPC2 magnitudes we used values of zeropoints and {\tt photflam} from the WFPC2 data handbook, along with filter wavelengths and bandwidths from the WFPC2 filters archive\footnote{\tt{http://www.stsci.edu/hst/wfpc2/documents/wfpc2\_filters\_\\archive.html}}.  For the ACS data, we used zeropoints from \cite{sirianni_etal_2005}, as used by \cite{gladstone_etal_2013}, appropriate {\tt photflam} and {\tt photplam} values for the observation dates calculated with the {\it HST} ACS zeropoint calculator\footnote{\tt{http://www.stsci.edu/hst/acs/analysis/zeropoints/zpt.py}}, and {\tt photbw} from the {\it HST} ACS data analysis web page\footnote{\tt{http://www.stsci.edu/hst/acs/analysis/bandwidths}}.  Having been converted into units of $\rm photons~cm^{-2}~s^{-1}$, {\sc xspec} readable spectra were created for each {\it HST} observation using the {\sc ftool} {\sc flx2xsp}.

\section{Data analysis}\label{data_analysis}

Having extracted the {\it XMM-Newton} EPIC spectra, and converted the {\it HST} magnitudes to an {\sc xspec} readable format, the goal of the analysis was to simultaneously fit the X-ray and optical/infrared data with the new {\sc optxirr} spectral model.  As several of the ULXs display some degree of X-ray flux variability, we decided to flux bin the X-ray observations. However, the ULXs were not observed contemporaneously in each of the bands, so we do not know which X-ray flux corresponds to the {\it HST} epochs; so as a compromise, the {{\it HST} data were simultaneously fitted with each flux bin of {\it XMM-Newton} data, and each possibility was assessed.  

To flux bin the X-ray data an approximate flux was extracted from each observation using an absorbed {\it p}-free disc model -- {\sc constant} $\times$ {\sc tbabs} $\times$ {\sc tbabs} $\times$ {\sc diskpbb} in {\sc xspec} 12.6.0.  The {\it p}-free disc model allows a range of values for the exponent of the radial dependence of temperature {\it p}, with $p = 0.75$ being a standard multi-colour-disc, and $p<0.75$ indicating advection; this allows for a broader spectrum, so is appropriate for approximating the flux in this ULX sample.  A multiplicative constant was included in the spectral model, which was frozen to unity for the pn detector (or the MOS1 detector when no pn data were available), and free to vary for the other EPIC detectors to allow for calibration uncertainties between the detectors \citep{read_etal_2014}.  {\sc tbabs} and the \cite{wilms_etal_2000} abundance table were used to model absorption in the X-ray spectrum; two such components were included, the first was fixed to account for the foreground Galactic column in the direction of the ULX, and the other left free to model any intrinsic absorption in the source and its host galaxy.  0.3--10 keV fluxes were estimated for the pn data (or MOS1 when no pn data was available) using the {\sc flux} command in {\sc xspec} and are shown in Table~\ref{xmm_obs_log}.  The {\it XMM-Newton} observations of each ULX were then grouped according to these fluxes: NGC 253 ULX2 and NGC 2403 X-1 varied by $\la 1 \times 10^{-12}~{\rm erg~cm^{-2}~s^{-1}}$, so a single flux group was used.  Two flux groups were used for NGC 1313 X-2, M81 X-6 and NGC 4190 ULX1, these were: $\sim$ 1--2 and $\sim$ (2--4) $\times 10^{-12}~{\rm erg~cm^{-2}~s^{-1}}$ for NGC 1313 X-2 (which agrees with the classification by \citealt{pintore_and_zampieri_2012}, based on spectral parameters); $\sim 1$ and $\sim$ (3--4) $\times 10^{-12}~{\rm erg~cm^{-2}~s^{-1}}$ for M81 X-6; $\sim 3$ and $\sim 6$ $\times 10^{-12}~{\rm erg~cm^{-2}~s^{-1}}$ for NGC 4190 ULX1.

The flux-grouped X-ray spectra were then fitted simultaneously with the {\it HST} data using {\sc xspec} and an absorbed, irradiated, colour-temperature-corrected disc -- {\sc constant} $\times$ {\sc tbabs} $\times$ {\sc tbabs} $\times$ {\sc redden} $\times$ {\sc optxirr}.  Again, a multiplicative constant was included to allow for calibration uncertainties between the EPIC detectors, and for slight variation in flux between the observations.  The constant was set equal to unity for the channel corresponding to the pn detector in the earliest observation in each flux group, and free for each of the other {\it XMM-Newton} data sets; whilst for the {\it HST} data, it was set to unity.  The first absorption component was fixed to the Galactic column in the direction of the source, whilst the second was left free to model any intrinsic absorption in the ULX or host galaxy.  The infrared/optical extinction (\sc redden}) was linked to the extra-galactic X-ray absorption assuming ${\rm E(B-V)} \approx 1.5 \times 10^{-22} N_{\rm H}$ \citep{gorenstein_1975}; Galactic extinction had already been corrected for in the magnitudes taken from \cite{gladstone_etal_2013}.  

The initial spectral fits with all variables free (except for the assumptions described in section \ref{optxirr}) were unable to place constraints on the fit parameters, so it was repeated with combinations of fixed black hole masses, outer disc radii, and ranges of the coronal radii.  Two characteristic black hole masses were used, these corresponded both to a typical 10 $\Msun$ sMBH and a 30 $\Msun$ MsBH \citep{feng_and_soria_2011}, and are consistent with the range of black hole masses thought to power the majority of ULXs (e.g. \citealt{feng_and_soria_2011}, and references therein). 

To test the validity of our assumed characteristic black hole masses, we checked whether an alternative explanation, i.e. IMBHs, could produce the observed broadband spectra.  We did this by attempting to fit the X-ray and optical data from the broadened disc ULXs using the absorbed/reddened {\sc optxirr} model with a 100 and 1000 $\Msun$ black hole.  As these black hole masses would require the ULXs to be at a sub-Eddington luminosity, it is appropriate to use sub-Eddington BHB-like spectra.  As such, in each case all of the power produced within the coronal radius was radiated as a power-law.  These models were rejected at greater than $3 \sigma$ significance in most of the sources, with the exceptions being the low flux X-ray data of M81 X-6, and a combination of a 100 $\Msun$ black hole and the low flux X-ray data of NGC 1313 X-2.  In the case of M81 X-6, the low flux X-ray data can be fitted with either black hole mass.  However, this is the lowest quality X-ray data in our sample, and such massive black holes can be rejected, as they are unable to reproduce the hard X-ray spectrum seen in this source at high flux.  For NGC 1313 X-2, the 100 $\Msun$ black hole model implies that the ULX has a bolometric luminosity of  $\sim 0.2 L_{\rm Edd}$, with a power-law spectral index of $\Gamma \sim 2.2$.  This is similar to the very high state in Galactic black holes (e.g. \citealt{mcclintock_and_remillard_2006}).  However, the observed high energy curvature \citep{stobbart_etal_2006} and spectral progression with X-ray luminosity \citep{feng_and_kaaret_2007} are inconsistent with a sub-Eddington state identification in this source; also, the reprocessing fraction required to produce the optical spectrum ($f_{\rm out} \ge 5$ per cent) is a factor of $\ga 10$ greater than would be expected for a sub-Eddington BHB (see Section~\ref{discussion} and \citealt{gierlinski_etal_2009}).  Hence we reject high black hole masses, and consider only 10 and 30 $\Msun$ black holes in our analyses.

The data were fitted with both an irradiated colour-temperature-corrected disc alone (i.e., with the coronal radius fixed to $6~r_{\rm g}$, and the electron temperature/optical depth frozen to arbitrary values, as the choice of coronal radius is such that there is no corona), and a disc plus Comptonising corona.  An outer disc radius of $10^5~r_{\rm g}$ (i.e. 1.5--4.5 $\times 10^{11}~{\rm cm}$ for a 10--30 $\Msun$ black hole) is consistent with limits on the size of the disc implied by broad emission lines in the ULX NGC 5408 X-1 \citep{cseh_etal_2011}.  However, larger outer disc radii of $\sim 10^{12}~{\rm cm}$ have been implied for a number of ULXs based on irradiated disc models (NGC 1313 X-1, \citealt{yang_etal_2011}; Ho II X-1, \citealt{tao_etal_2012}; Ho IX X-1, \citealt{tao_etal_2011}; NGC 5408, \citealt{grise_etal_2012}).  In this case, an outer disc radius of $\sim 10^{6}~r_{\rm g}$ is appropriate ($10^{6}~{r_{\rm g}} \approx$ (1.5--4.5) $\times 10^{12}~{\rm cm}$ for a 10--30 $\Msun$ black hole).  So, the ULX spectra were fitted using both fixed outer disc radii of $10^5$ and $10^6~r_{\rm g}$.

\section{Results}

\begin{figure*}
\begin{center}
\includegraphics[width=15cm]{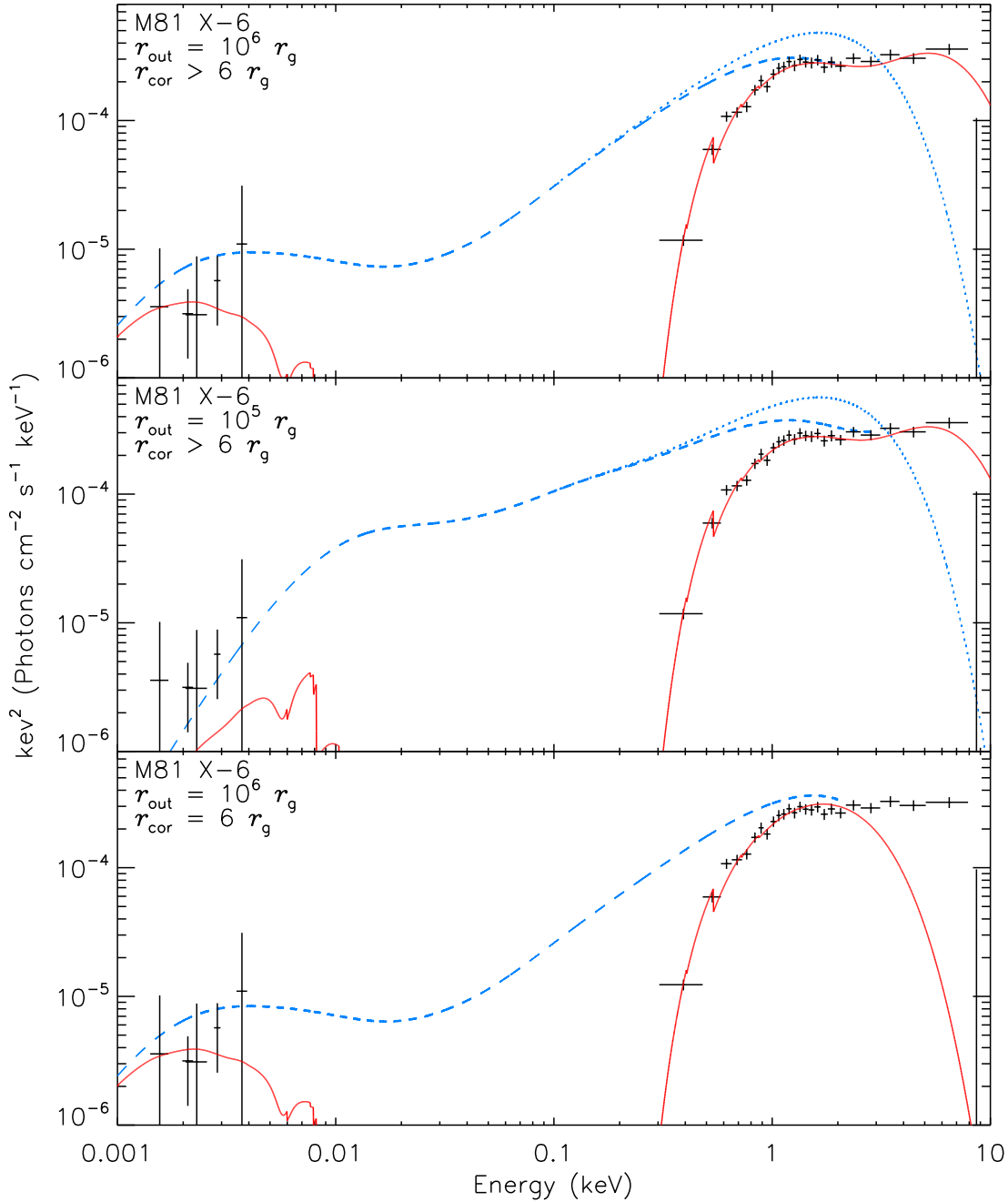}
\caption{Combined optical and X-ray spectrum of M81 X-6, showing both the {\it HST} and low flux {\it XMM-Newton} data, overplotted with several different models.  For clarity, only the pn data from observation 0657802201 is shown, and it is rebinned to 10 $\sigma$ significance.  The solid red line shows the observed model (the model and data are corrected for Galactic reddening at optical wavelengths as described in the text, but this has only a small effect), and the dashed blue line shows the deabsorbed model.  Where it is shown, the dotted blue line corresponds to the intrinsic disc spectrum, if the inner corona is removed from the model.  ({\it top}) The data overplotted with a model of a 30 $\Msun$ black hole with an outer disc radius of $10^6~{r_{\rm g}}$ and reproccessing of the inner disc emission in a corona.  This is the statistically best fitting model, and results in a value of $f_{\rm out} \sim 0.5$ per cent.  ({\it centre}) This model corresponds to a 30 $\Msun$ black hole with an outer disc radius of $10^5~{r_{\rm g}}$ and a central corona.  The choice of a smaller outer disc radius results in poor constraints on the irradiating fraction ($\le 40$ per cent).  ({\it bottom}) A 30 $\Msun$ black hole with an outer disc radius of $10^6~{r_{\rm g}}$, but no corona.  Note that this model is not a statistically acceptable fit.}
\label{spectra}
\end{center}
\end{figure*}

The results of the spectral fitting are shown in Tables \ref{results1} and \ref{results2}, and examples of the spectral models are shown in Figure~\ref{spectra} for the optical and low X-ray flux data from M81 X-6.  Formally acceptable spectral fits (not rejected at $3 \sigma$ significance) were found for each of the 5 ULXs, although the best fit to the NGC 253 ULX2 data could be rejected at greater than the 99 per cent significance level.  For the 3 ULXs with 2 flux bins of X-ray data, the best fitting model was found for the high flux data in NGC 1313 X-2, and the low flux data in M81 X-6 and NGC 4190 ULX1.  However, acceptable models were found for both sets of flux binned X-ray data sets in NGC 1313 X-2 and NGC 4190 ULX1.  As we cannot distinguish which (or indeed, if any) of the flux binned X-ray data corresponds to each epoch of optical data, we continue to consider both possibilities.  Also, it should be noted that due to the relative amounts of {\it XMM-Newton} and {\it HST} data, the $\chi^2$ statistic is heavily dominated by the X-ray data.  Therefore, the lack of an acceptable fit for the optical and high flux M81 X-6 X-ray data does not allow us to make conclusions about the unobserved X-ray spectrum during the optical observations, rather it mainly indicates that the model was unable to reproduce the high flux X-ray spectrum.

\begin{table*}
\centering
\caption{{\sc optxirr} fits with outer disc radius fixed at $10^{5}~r_{\rm g}$}
\begin{tabular}{cccccccc}
\hline
$N_{\rm H}$$^a$ & Mass$^b$ & log($L_{\rm bol}/L_{\rm Edd}$)$^c$ & $r_{\rm cor}$$^d$ & $kT_{\rm e}$$^e$ & $\tau$$^f$ & $f_{\rm out}$$^g$ & $\chi^2/{\rm d.o.f.}$$^h$ \\
\hline
\multicolumn{8}{c}{{\bf NGC 253 ULX2}} \\
$0.236 \pm 0.003$ & 10 & $0.297 \pm 0.006$ & 6 & - & - & [$3 \times 10^{-3}$] & [4865.6/3092] \\
$0.206 \pm 0.004$ & 10 & $0.200^{+0.008}_{-0.009}$ & $62 \pm 8$ & $1.26 \pm 0.02$ & $14.9^{+0.5}_{-0.4}$ & [$3 \times 10^{-3}$] & {\bf 3307.7/3089} \\
{[}0.4] & 30 & [-0.1] & 6 & - & - & [$2 \times 10^{-2}$] & [13593.2/3092] \\
$0.296^{+0.004}_{-0.002}$ & 30 & $-0.241^{+0.008}_{-0.009}$ & $\ge 100$ & $1.19 \pm 0.01$ & $14.7 \pm 0.2$ & $\ge 1 \times 10^{-5}$ & [3394.8/3089] \\\\

\multicolumn{8}{c}{{\bf NGC 1313 X-2}} \\
\multicolumn{8}{c}{\it high X-ray flux} \\
$0.122 \pm 0.002$ & 10 & $0.512 \pm 0.007$ & 6 & - & - & $1.1^{+0.8}_{-0.5} \times 10^{-2}$ & [6009.4/4504] \\
$0.158 \pm 0.003$ & 10 & $0.30 \pm 0.02$ & $27 \pm 2$ & $1.49 \pm 0.04$ & $20^{+7}_{-3}$ & [$4 \times 10^{-3}$] & {\bf 4534.3/4501} \\
{[}0.2] & 30 & [0.2] & 6 & - & - & [$1 \times 10^{-2}$] & [16307.6/4504] \\
$0.246^{+0.002}_{-0.003}$ & 30 & $-0.11^{+0.01}_{-0.02}$ & $\ge 100$ & $1.39^{+0.02}_{-0.01}$ & $12.0 \pm 0.1$ & $3^{+12}_{-2} \times 10^{-2}$ & 4666.7/4501 \\\\
\multicolumn{8}{c}{\it low X-ray flux} \\
$(9 \pm 2) \times 10^{-3}$ & 10 & $-0.163 \pm 0.007$ & 6 & - & - & [$9 \times 10^{3}$] & [2544.2/1457] \\
$8.5^{+0.4}_{-0.3} \times 10^{-2}$ & 10 & $0.29 \pm 0.02$ & $73 \pm 3$ & $\ge 7$ & $4.3 \pm 0.1$ & [$3 \times 10^{-3}$] & [1790.7/1454] \\
{[}0.1] & 30 & [-0.6] & 6 & - & - & [$7 \times 10^{-2}$] & [3303.4/1457] \\
$0.130^{+0.002}_{-0.004}$ & 30 & $-0.46^{+0.02}_{-0.01}$ & $\ge 100$ & $2.6^{+0.6}_{-0.4}$ & $5.3^{+0.6}_{-0.8}$ & $6^{+13}_{-4} \times 10^{-2}$ & 1602.4/1454 \\\\

\multicolumn{8}{c}{{\bf NGC 2403 X-1}} \\
$0.155^{+0.004}_{-0.005}$ & 10 & $-0.12 \pm 0.01$ & 6 & - & - & [$8 \times 10^{-3}$] & 1136.1/1009 \\
$0.161 \pm 0.004$ & 10 & $-0.16 \pm 0.02$ & $11.6 \pm 0.6$ & $1.08^{+0.09}_{-0.06}$ & $\ge 30$ & [$1 \times 10^{-2}$] & 1102.9/1006 \\
{[}0.3] & 30 & [-0.4] & 6 & - & - & [$5 \times 10^{-2}$] & [2762.9/1009] \\
$0.252 \pm 0.007$ & 30 & $-0.56 \pm 0.02$ & $40 \pm 10$ & $0.90 \pm 0.02$ & $16^{+3}_{-1}$ & $7^{+51}_{-6} \times 10^{-2}$ & {\bf 1071.6/1006} \\\\

\multicolumn{8}{c}{{\bf M81 X-6}} \\
\multicolumn{8}{c}{\it high X-ray flux} \\
$(4.2 \pm 0.2) \times 10^{-2}$ & 10 & $0.401 \pm 0.003$ & 6 & - & - & [$2 \times 10^{-3}$] & [2969.1/2436] \\
$0.111 \pm 0.002$ & 10 & $0.417 \pm 0.005$ & $10.6^{+0.4}_{-0.1}$ & $1.8 \pm 0.1$ & $\ge 30$ & [$2 \times 10^{-3}$] & [2905.8/2433] \\
{[}0.3] & 30 & [-0.8] & 6 & - & - & [$5 \times 10^{-2}$] & [8825.2/2436] \\
$0.200 \pm 0.003$ & 30 & $-4.5^{+0.3}_{-0.5} \times 10^{-2}$ & $62^{+8}_{-25}$ & $1.11 \pm 0.01$ & $15.0^{+0.7}_{-0.6}$ & $\le 0.2$ & [2801.1/2433] \\\\
\multicolumn{8}{c}{\it low X-ray flux} \\
$(4.0 \pm 0.6) \times 10^{-2}$ & 10 & $-0.12 \pm 0.02$ & 6 & - & - & [$1 \times 10^{-2}$] & [485.1/279] \\
$0.14 \pm 0.01$ & 10 & $0.35 \pm 0.03$ & $\ge 100$ & $\ge 9$ & $4.1^{+0.5}_{-0.2}$ & [$2 \times 10^{-3}$] & 281.6/276 \\
{[}0.1] & 30 & [-0.6] & 6 & - & - & [$7 \times 10^{-2}$] & [680.2/279] \\
$0.164^{+0.009}_{-0.008}$ & 30 & $-0.43 \pm 0.02$ & $25^{+10}_{-3}$ & $1.4^{+0.3}_{-0.1}$ & $\ge 10$ & $\le 0.4$ & {\bf 262.1/276} \\\\

\multicolumn{8}{c}{{\bf NGC 4190 ULX1}} \\
\multicolumn{8}{c}{\it high X-ray flux} \\
$(5.8 \pm 0.3) \times 10^{-2}$ & 10 & $0.446 \pm 0.003$ & 6 & - & - & $3^{+15}_{-2} \times 10^{-3}$ & [1163.6/920] \\
$(6.5 \pm 0.3) \times 10^{-2}$ & 10 & $0.494 \pm 0.005$ & $19.1^{+2.8}_{-0.5}$ & $1.56^{+0.08}_{-0.03}$ & $\ge 40$ & $3^{+14}_{-2} \times 10^{-3}$ & 924.3/917 \\
{[}0.2] & 30 & [$-8 \times 10^{-2}$] & 6 & - & - & [$4 \times 10^{-2}$] & [3548.4/920] \\
$0.140 \pm 0.005$ & 30 & $2.5 \pm 0.5 \times 10^{-2}$ & $\ge 90$ & $1.35 \pm 0.03$ & $12.8^{+0.6}_{-0.2}$ & $8^{+13}_{-4} \times 10^{-3}$ & 963.6/917 \\\\
\multicolumn{8}{c}{\it low X-ray flux} \\
$4.9 \pm 0.3 \times 10^{-2}$ & 10 & $0.23 \pm 0.01$ & 6 & - & - & $6^{+23}_{-4} \times 10^{-3}$ & 863.8/845 \\
$5.2 \pm 0.3 \times 10^{-2}$ & 10 & $0.190^{+0.003}_{-0.014}$ & $13.9^{+0.9}_{-0.6}$ & $1.15^{+0.03}_{-0.04}$ & $\ge 40$ & $6^{+25}_{-4} \times 10^{-3}$ & {\bf 803.8/842} \\
{[}0.2] & 30 & [-0.1] & 6 & - & - & [$3 \times 10^{-2}$] & [2488.6/845] \\
$0.121^{+0.006}_{-0.005}$ & 30 & $-0.25 \pm 0.02$ & $\ge 60$ & $1.03 \pm 0.03$ & $16 \pm 1$ & $1.0^{+0.9}_{-0.5} \times 10^{-2}$ & 806.1/842 \\

\hline
\end{tabular}
\label{results1}
\begin{minipage}{\linewidth}
Spectral parameters from {\sc tbabs} $\times$ {\sc tbabs} $\times$ {\sc redden} $\times$ {\sc optxirr}, with an outer radius of $10^{5}~r_{\rm g}$.  Errors and limits indicate the $1 \sigma$ confidence regions.  Parameters in square brackets are not constrained at the $1 \sigma$ level by the model.  Based on the fit statistic, we reject a 30 $\Msun$ black hole with no corona in all cases, and a 10 $\Msun$ black hole with no corona in 6 out of 8 cases.
$^a$Extra-galactic absorption column ($\times 10^{22}~{\rm cm^{-2}}$).
$^b$Black hole mass ($\Msun$), this was fixed to 2 characteristic values - 10 and 30 $\Msun$.
$^c$Logarithm of the Eddington ratio.
$^d$Outer coronal radius ($r_{\rm g}$), this is fixed to $6~r_{\rm g}$ in models with no corona.
$^e$Electron temperature of the Comptonised corona (keV).
$^f$Optical depth of the Comptonised corona.
$^g$Fraction of the flux from the central regions of the source that is  reprocessed in the outer disc.
$^h$Goodness of fit in terms of $\chi^2/$degrees of freedom, values in bold indicate the statistical best fit for each source, whilst values in square brackets indicate fits which are rejected at $3 \sigma$ significance.
\end{minipage}
\end{table*}

\begin{table*}
\centering
\caption{{\sc optxirr} fits with outer disc radius fixed at $10^{6}~r_{\rm g}$}
\begin{tabular}{cccccccc}
\hline
$N_{\rm H}$$^a$ & Mass$^b$ & log($L_{\rm bol}/L_{\rm Edd}$)$^c$ & $r_{\rm cor}$$^d$ & $kT_{\rm e}$$^e$ & $\tau$$^f$ & $f_{\rm out}$$^g$ & $\chi^2/{\rm d.o.f.}$$^h$ \\
\hline
\multicolumn{8}{c}{{\bf NGC 253 ULX2}} \\
$0.236 \pm 0.003$ & 10 & $0.297 \pm 0.006$ & 6 & - & - & $0.3 \pm 0.2$& [4848.4/3092] \\
$0.206 \pm 0.004$ & 10 & $0.199 \pm 0.007$ & $62^{+10}_{-3}$ & $1.26^{+0.02}_{-0.01}$ & $15.0^{+0.5}_{-0.4}$ & $0.4^{+0.3}_{-0.2}$ & {\bf 3288.7/3089} \\
{[}0.4] & 30 & [-0.1] & 6 & - & - & [1] & [13545.4/3092] \\
$0.296 \pm 0.004$ & 30 & $-0.241 \pm 0.009$ & $\ge 100$ & $1.19 \pm 0.01$ & $14.7 \pm 0.2$ & $\ge 0.6$ & [3340.3/3089] \\\\

\multicolumn{8}{c}{{\bf NGC 1313 X-2}} \\
\multicolumn{8}{c}{\it high X-ray flux} \\
$0.122 \pm 0.002$ & 10 & $0.512 \pm 0.007$ & 6 & - & - & $1.1^{+0.8}_{-0.5} \times 10^{-2}$ & [6009.4/4504] \\
$0.158 \pm 0.003$ & 10 & $0.31 \pm 0.02$ & $27^{+3}_{-2}$ & $1.48 \pm 0.04$ & $20^{+8}_{-3}$ & $3^{+2}_{-1} \times 10^{-2}$ & {\bf 4518.8/4501} \\
{[}0.2] & 30 & [0.2] & 6 & - & - & [$2 \times 10^{-3}$] & [16300.3/4504] \\
$0.246 \pm 0.003$ & 30 & $-0.11^{+0.01}_{-0.02}$ & $\ge 100$ & $1.39^{+0.02}_{-0.01}$ & $12.0 \pm 0.1$ & $(4 \pm 2) \times 10^{-3}$ & 4660.0/4501 \\\\
\multicolumn{8}{c}{\it low X-ray flux} \\
$(9 \pm 2) \times 10^{-3}$ & 10 & $-0.163 \pm 0.007$ & 6 & - & - & $1.9^{+1.0}_{-0.7} \times 10^{-2}$ & [2528.7/1457] \\
$(3.8 \pm 0.3) \times 10^{-2}$ & 10 & $-0.08 \pm 0.02$ & $\ge 100$  & $3.1^{+0.6}_{-0.8}$ & $4.7^{+0.6}_{-0.8}$ & $2.0^{+1.1}_{-0.8} \times 10^{-2}$ & [1774.9/1454] \\
{[}0.1] & 30 & [-0.6] & 6 & - & - & [$7 \times 10^{-3}$] & [3297.6/1457] \\
$0.129 \pm 0.003$ & 30 & $-0.46^{+0.02}_{-0.01}$ & $\ge 100$ & $2.6^{+0.7}_{-0.4}$ & $5.3^{+0.6}_{-0.7}$ & $6^{+3}_{-2} \times 10^{-3}$ & 1595.9/1454 \\\\

\multicolumn{8}{c}{{\bf NGC 2403 X-1}} \\
$0.155^{+0.005}_{-0.004}$ & 10 & $-0.11 \pm 0.01$ & 6 & - & - & $(3 \pm 2) \times 10^{-3}$ & 1132.3/1009 \\
$0.161 \pm 0.005$ & 10 & $-0.16 \pm 0.02$ & $10.9^{+0.7}_{-0.6}$ & $1.09^{+0.08}_{-0.07}$ & $\ge 30$ & $4^{+3}_{-2} \times 10^{-3}$ & 1099.1/1006 \\
{[}0.3] & 30 & [-0.4] & 6 & - & - & [$2 \times 10^{-3}$] & [2761.9/1009] \\
$0.253^{+0.008}_{-0.007}$ & 30 & $-0.56 \pm 0.02$ & $40 \pm 10$  & $0.90 \pm 0.02$ & $16^{+3}_{-2}$ & $3^{+2}_{-1} \times 10^{-3}$ & {\bf 1071.1/1006} \\\\

\multicolumn{8}{c}{{\bf M81 X-6}} \\
\multicolumn{8}{c}{\it high X-ray flux} \\
$0.107 \pm 0.002$ & 10 & $0.401 \pm 0.003$ & 6 & - & - & $5^{+4}_{-3} \times 10^{-3}$ & [2962.4/2436]$^i$ \\
{[}0.3] & 30 & [$-8 \times 10^{-2}$] & 6 & - & - & [$4 \times 10^{-3}$] & [8820.1/2436] \\
$0.200 \pm 0.003$ & 30 & $(-4.5 \pm 0.3) \times 10^{-2}$ & $62^{+8}_{-7}$  & $1.11^{+0.03}_{-0.01}$ & $15.0^{+0.7}_{-0.6}$ & $(3 \pm 1) \times 10^{-3}$ & [2797.5/2433] \\\\
\multicolumn{8}{c}{\it low X-ray flux} \\
$(4.0 \pm 0.6) \times 10^{-2}$ & 10 & $-0.12 \pm 0.02$ & 6 & - & - & $1.0^{+0.7}_{-0.5} \times 10^{-2}$ & [478.6/279] \\
$0.14 \pm 0.01$ & 10 & $0.36 \pm 0.04$ & $\ge 100$ & $\ge9$ & $4.1^{+0.4}_{-0.3}$ & $7^{+7}_{-4} \times 10^{-3}$ & 274.9/276 \\
{[}0.1] & 30 & [-0.6] & 6 & - & - & [$6 \times 10^{-3}$] & [677.2/279] \\
$0.165^{+0.010}_{-0.009}$ & 30 & $-0.43 \pm 0.02$ & $25^{+10}_{-4}$ & $1.4^{+0.4}_{-0.1}$ & $\ge 11$ & $5^{+3}_{-2} \times 10^{-3}$ & {\bf 258.7/276} \\\\

\multicolumn{8}{c}{{\bf NGC 4190 ULX1}} \\
\multicolumn{8}{c}{\it high X-ray flux} \\
$(5.8 \pm 0.3) \times 10^{-2}$ & 10 & $0.446 \pm 0.005$ & 6 & - & - & $(8 \pm 3) \times 10^{-4}$ & [1157.5/920] \\
$(6.5 \pm 0.4) \times 10^{-2}$ & 10 & $0.494^{+0.004}_{-0.007}$ & $19.0^{+4.0}_{-0.6}$ & $1.56^{+0.13}_{-0.05}$ & $\ge 17$ & $(8 \pm 3) \times 10^{-4}$ & 917.9/917 \\
{[}0.2] & 30 & [$-8 \times 10^{-2}$] & 6 & - & - & [$1 \times 10^{-3}$] & [3555.6/920] \\
$0.139 \pm 0.005$ & 30 & $2.5^{+0.6}_{-0.5} \times 10^{-2}$ & $\ge 90$  & $1.3^{+0.6}_{-0.2}$ & $12.9^{+0.6}_{-0.2}$ & $(7 \pm 4) \times 10^{-4}$ & 969.1/917 \\\\
\multicolumn{8}{c}{\it low X-ray flux} \\
$(4.9 \pm 0.4) \times 10^{-2}$ & 10 & $0.23 \pm 0.01$ & 6 &  - & - & $1.3^{+0.5}_{-0.4} \times 10^{-3}$ & 858.2/845 \\
$(5.2 \pm 0.4) \times 10^{-2}$ & 10 & $0.19 \pm 0.02$ & $14.0^{+1.6}_{-0.7}$ & $1.15^{+0.06}_{-0.05}$ & $\ge 20$ & $(1.5 \pm 0.5) \times 10^{-3}$ & {\bf 798.1/842} \\
{[}0.2] & 30 & [-0.1] & 6 & - & - & [$1 \times 10^{-3}$] & [2495.4/845] \\
$0.121^{+0.006}_{-0.005}$ & 30 & $-0.25 \pm 0.02$ & $\ge 60$  & $1.03 \pm 0.03$ & $16 \pm 1$ & $1.3^{+0.7}_{-0.6} \times 10^{-3}$ & 811.2/842 \\

\hline
\end{tabular}
\label{results2}
\begin{minipage}{\linewidth}
Spectral parameters from {\sc tbabs} $\times$ {\sc tbabs} $\times$ {\sc redden} $\times$ {\sc optxirr}, with an outer radius of $10^{6}~r_{\rm g}$.  Errors and limits indicate the $1 \sigma$ confidence regions.  Parameters in square brackets are not constrained at the $1 \sigma$ level by the model.  Similarly to the fits reported in Table \ref{results1}, based on the fit statistic, we reject a 30 $\Msun$ black hole with no corona in all cases, and a 10 $\Msun$ black hole with no corona in 6 out of 8 cases.  Typically, these models with an outer disc radius of $10^{6}~r_{\rm g}$ are better constrained than the previous case, with an outer radius of $10^{5}~r_{\rm g}$, and they have improved fit statistics (albeit marginal in most cases). 
$^a$Extra-galactic absorption column ($\times 10^{22}~{\rm cm^{-2}}$).
$^b$Black hole mass ($\Msun$), this was fixed to 2 characteristic values - 10 and 30 $\Msun$.
$^c$Logarithm of the Eddington ratio.
$^d$Outer coronal radius ($r_{\rm g}$), this is fixed to $6~r_{\rm g}$ in models with no corona.
$^e$Electron temperature of the Comptonised corona (keV).
$^f$Optical depth of the Comptonised corona.
$^g$Fraction of the flux from the central regions of the source that is  reprocessed in the outer disc.
$^h$Goodness of fit in terms of $\chi^2/$degrees of freedom, values in bold indicate the statistical best fit for each source, whilst values in square brackets indicate fits which are rejected at $3 \sigma$ significance.
$^i$No fit parameters corresponding to a 10 $\Msun$ black hole and a free corona are shown for the high flux X-ray data of M81 X-6, as this did not result in any improvement over a disc alone.
\end{minipage}
\end{table*}

In general, the inclusion of a Comptonised corona resulted in an improved fit to the data over a disc alone; furthermore, in three of the ULXs (NGC 253 ULX2, NGC 1313 X-2 and M81 X-6) all spectral models without a corona are rejected at $> 3 \sigma$ significance.  Perhaps unsurprisingly given that these sources were identified as having broadened disc-like spectra in \citep{sutton_etal_2013b}, a standard disc alone (with spin set to zero) is insufficient to reproduce the observed X-ray data.  Similarly to other ULX X-ray spectral studies (e.g. \citealt{dewangan_etal_2006}; \citealt{goad_etal_2006}; \citealt{stobbart_etal_2006}; \citealt{gladstone_etal_2009}), we found that the Comptonised coronae were generally cool and optically thick: the coronal parameters are typically constrained to $kT_{\rm e} \sim$  1--3 keV and $\tau \ga 10$.  There is no consistently favoured black hole mass - a 10 $\Msun$ black hole provides the best fit for NGC 253 ULX2, NGC 1313 X-2 and NGC 4190 ULX1, whilst a 30 $\Msun$ black hole is preferred for NGC 2403 X-1 and M81 X-6; although the only source where either of the black hole masses is formally rejected for all combinations of outer disc radius and coronal radius is NGC 253 ULX2, where a 30 $\Msun$ black hole cannot adequately describe the observed spectra.

\begin{figure*}
\begin{center}
\includegraphics[width=18cm]{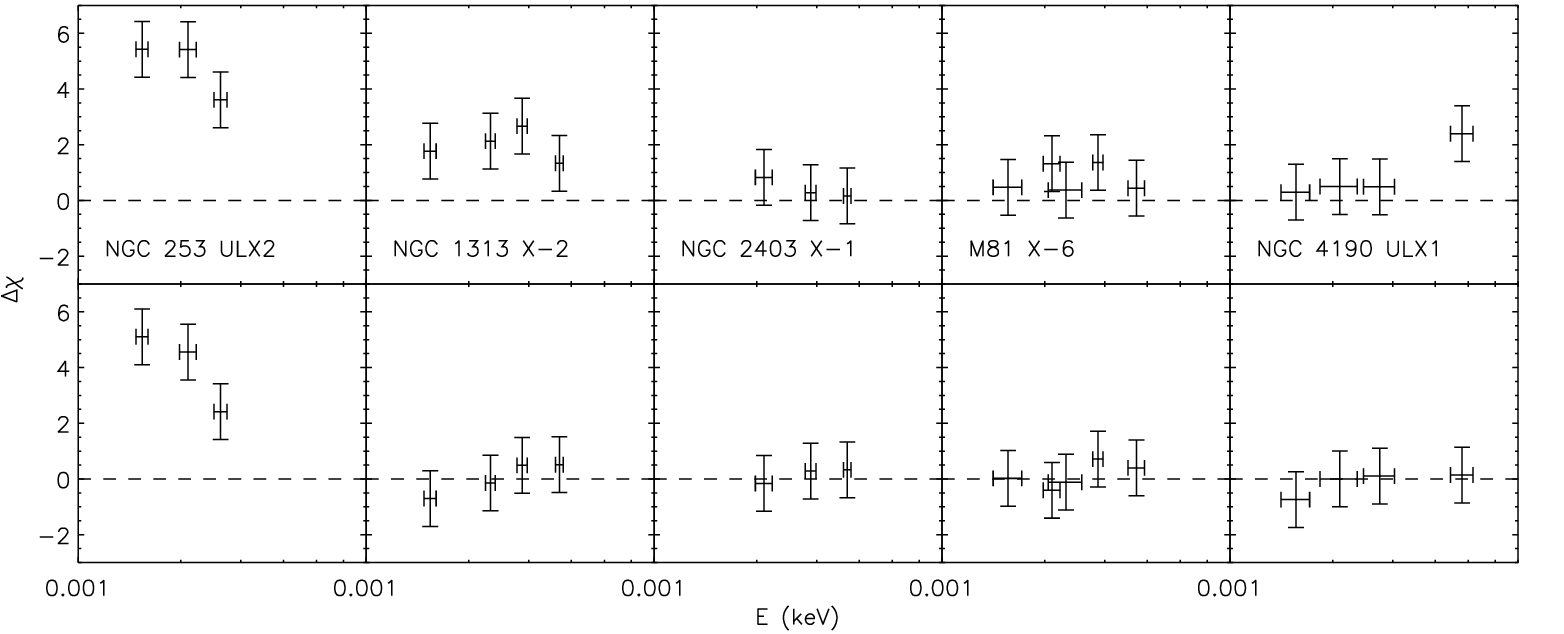}
\caption{{\it HST} residuals from the statistically best-fitting spectral model to each of the sample ULXs which are labelled from {\it left} to {\it right}, with model outer disc radii of $10^5~r_{\rm g}$ ({\it top} row) and $10^6~r_{\rm g}$ ({\it bottom} row), as shown in Tables~\ref{results1} and \ref{results2}.  The residuals are plotted in terms of $\Delta \chi$, and the dashed line corresponds to $\Delta \chi = 0$.}
\label{residuals1}
\end{center}
\end{figure*}

Spectral fitting was carried out for two fixed characteristic outer disc radii of $10^5$ and $10^6~r_{\rm g}$.  Typically, the absorption, Eddington ratio, coronal radius, electron temperature and optical depth vary little between models with the same black hole mass but different outer disc radii.  There are however slight reductions in the value of $\chi^2$ as the outer disc radius is increased.  For example, for the best fitting models of NGC 253 ULX2 and NGC 1313 X-1, $\Delta \chi^2 =$ 19 and 15.5 respectively.  The effect of varying the outer disc radius is clear from inspecting the spectral shape at optical wavelengths (Figure~\ref{spectra} {\it top} and {\it centre}), or the difference in the residuals to the spectral fit for each outer disc radius (Figure \ref{residuals1}).  The outer disc radius has little effect on the X-ray spectrum, but in combination with the fraction of emission reprocessed in the outer disc, it determines the contribution of the accretion disc to the optical spectrum (see also \citealt{gierlinski_etal_2009}).  From Figure \ref{residuals1}, it is clear that none of the spectral models can adequately reproduce the observed optical counterpart in NGC 253 ULX2.  Also, in NGC 1313 X-2 the $10^6~r_{\rm g}$ outer disc radius is an improvement over $10^5~r_{\rm g}$, which under-predicts each of the optical magnitudes by greater than the $1 \sigma$ confidence ranges.  In the other ULXs the difference between the models is more subtle, but the $1 \sigma$ confidence regions of the F435W and F606W fluxes in M81 X-6, and the F300W magnitude in NGC 4190 ULX1 lie outside of the model predictions for $r_{\rm out} = 10^5~r_{\rm g}$, and for these sources the optical data points which do lie within the $1 \sigma$ error regions of the model are particularly unconstrained, having $1 \sigma$ uncertainties of $\pm$ 2--3 magnitudes.

We also consider the fraction of emission reprocessed in the outer disc.  For the models with $r_{\rm out} = 10^5~r_{\rm g}$, $f_{\rm out}$ was very poorly constrained.  Where errors could be estimated, they were large, or only provided ineffectively high upper limits.  For the models with  $r_{\rm out} = 10^6~r_{\rm g}$, which the model residuals would argue provides a superior fit to the optical data, we were able to place better constraints on $f_{\rm out}$.  For NGC 253 ULX2, $f_{\rm out} \sim 0.4$, which seems unphysically large, however in this source it is clear from the fit residuals (Figure \ref{residuals1}) that the model is unable to reproduce the observed optical data.  For the other 4 ULXs, estimates of $f_{\rm out}$ range from $\sim$ 0.1--3 per cent.  Where there is both high and low flux X-ray data, as would be expected, the low flux X-ray data requires a larger fraction to be reprocessed.  If we consider only the statistically best fitting model for each ULX, then we estimate $f_{\rm out} = 3^{+2}_{-1} \times 10^{-2}$, $3^{+2}_{-1} \times 10^{-3}$, $5^{+3}_{-2} \times 10^{-3}$ and $(1.5 \pm 0.5) \times 10^{-3}$ for NGC 1313 X-2, NGC 2403 X-1, M81 X-6 and NGC 4190 ULX respectively.  With the exception of NGC 1313 X-2, these are all in the range of a few $\times 10^{-3}$; furthermore, the reprocessed fraction quoted for NGC 1313 X-2 corresponds to a 10 $\Msun$ black hole, and it could be a factor of $\sim 10$ lower if it instead contains a MsBH ($f_{\rm out} \sim (4$--$6) \times 10^{-3}$ for a 30 $\Msun$ black hole).

Throughout the spectral fitting we have assumed a black hole spin of 0, and a model normalisation of 2 (i.e. the binary system is close to face on).  Here we test these assumptions, by taking the best fitting model for each source (excluding NGC 253 ULX2; see below), altering either the normalisation or spin, and refitting the model to the data.  We do this test for a normalisation of 1, and spin values of 0.5 and 0.998.  The model with a normalisation of 1 is statistically rejected (at $> 3 \sigma$ significance) in NGC 1313 X-2, and results in a worse fit in NGC 4190 ULX1 ($\Delta \chi^2 = -48.2$).  In NGC 2403 X-1 and M81 X-6, this model results in similar fit statistics to the corresponding model with a normalisation of 2 ($\Delta \chi^2 = 3.1$ and -6.8 in NGC 2403 X-1 and M81 X-6 respectively).  In these two cases, the decrease in normalisation is largely compensated for by a corresponding increase in $L_{\rm bol}/L_{\rm Edd}$.  Using a spin of 0.5 results in a worse fit in NGC 1313 X-2 ($\Delta \chi^2 = -113.4$), but similar fit statistics to the spin zero model in the other sources ($\Delta \chi^2 = 3.1$, -8.3 and 7.1 in NGC 2403 X-1, M81 X-6 and NGC 4190 ULX1 respectively).  In general, there are no significant changes to the parameters between the spin 0 and 0.5 models, although the higher spin results in a lower intrinsic absorption and coronal radius in NGC 1313 X-2, and a cooler corona in NGC 4190 ULX1.  The models with spins of 0.998 are rejected (at $> 3 \sigma$ significance) in NGC 1313 X-2 and M81 X-6.  Acceptable fits are found for NGC 2403 X-1 and NGC 4190 ULX1, although for NGC 4190 ULX1 the fit is worse than the spin zero model ($\Delta \chi^2 = -67.6)$, whilst a slight improvement is found for NGC 2403 X-1 ($\Delta \chi^2 = 3.1$).  Where we find acceptable fits, the parameters are mostly consistent with those from the spin zero model, except $L_{\rm Edd}/L_{\rm bol}$ is lower, the intrinsic absorption column in NGC 2403 X-1 is lower, and the corona is cooler in NGC 4190 ULX1.  Crucially, the irradiating fraction does not change significantly in any of the statistically acceptable models considered here.

\section{Discussion}\label{discussion}

In this study we have used a new model of black hole accretion, which includes both a colour-temperature-correction and irradiation of the outer accretion disc to model the X-ray and optical spectra of a sample of ULXs.  These have both {\it HST} and {\it XMM-Newton} archival data, show broadened disc-like X-ray spectra in \cite{sutton_etal_2013b} and have optical counterparts identified by \cite{gladstone_etal_2013}. 

Using the {\sc optxirr} model, we were able to reproduce the observed spectra in 4 of the 5 sample ULXs; for NGC 253 ULX2 it is evident from the fit residuals that the model is unable to produce the observed optical flux.  It is noteworthy that the {\it HST} counterpart to NGC 253 ULX2 is significantly brighter than most other known ULX counterparts in this sample; at a distance of 3.68 Mpc, NGC 253 ULX2 has a distance modulus of 27.8, thus the absolute magnitudes (Galactic-reddening-corrected absolute magnitudes of -6.1, -7.0 and -8.2 in the F475W, F606W and F814W bands respectively) and colours (${\rm F475W - F606W} \approx 0.9$ and ${\rm F475W - F814W} \approx 2.1$; corrected for Galactic reddening only) of the counterpart may be suggestive of a stellar cluster in NGC 253.  As the identity of the counterpart is uncertain, we do not consider NGC 253 ULX2 further here.

Of the 4 other ULXs in the sample, 3 -- NGC 1313 X-2, M81 X-6 and NGC 4190 ULX1 -- had been observed to display a degree of inter-observational variability in X-ray flux; the X-ray spectra were flux-grouped, but with the lack of simultaneous {\it XMM-Newton} and {\it HST} observations, there is no way to determine which of the X-ray flux groups corresponds to the epoch of the optical observations.  For NGC 1313 X-2 and NGC 4190 ULX1 we find statistically acceptable spectral fits at the $3 \sigma$ level for both sets of flux grouped X-ray data, although the nominal best fits corresponded to the high flux and low flux X-ray spectra respectively.  However, in the case of M81 X-6, no acceptable spectral fit was identified for the high flux X-ray data, indicating that the data were not consistent with our X-ray model.  For 2 sources -- NGC 1313 X-2 and M81 X-6 -- an irradiated, colour-temperature-corrected disc alone could not adequately fit the observed data; and, even in NGC 2403 X-1 and NGC 4190 ULX1, where a disc alone could not be rejected, the addition of a Comptonised component improved the fit statistic ($\Delta \chi^2$ = 33.2 and 60.1 for 3 degrees of freedom, between the best fitting disc and the equivalent model with the addition of a Compton corona).  This is entirely consistent with our identification of these X-ray spectra as broadened discs.  Also, neither of the characteristic black hole masses could be rejected in any of the ULXs, although a 10 $\Msun$ black hole provided the preferred solution for NGC 1313 X-2 and NGC 4190 ULX1 with mildly super-Eddington luminosities ($L_{\rm bol}/L_{\rm Edd} \sim$ 2), and a 30 $\Msun$ black hole for NGC 2403 X-1 and M81 X-6 with sub-Eddington luminosities ($L_{\rm bol}/L_{\rm Edd} \sim 0.3$ and 0.4).  Such sub-Eddington luminosities are arguably rather unusual, given the seemingly ubiquitous hard, cool Compton corona suggested by the X-ray data, which is unlike any of the known sub-Eddington accretion states.  We discuss this further below. 

Of the two characteristic outer disc radii tested, neither can be ruled out based on the $\chi^2$ statistic in any of the sample sources.  But, it is evident from a comparison of Tables \ref{results1} and \ref{results2} that the $10^6~r_{\rm g}$ outer disc radius offered at least a marginally improved fit in most of the models.  However, given the relative amounts of optical and X-ray data points used in this analysis, the $\chi^2$ statistic is heavily dominated by the X-ray spectrum.  We demonstrate this by taking the low X-ray flux M81 X-6 data, changing the {\it HST} fluxes in increments of 1 magnitude, and re-fitting the model for a 30 {\Msun} black hole with a Comptonising corona.  Increasing or decreasing the {\it HST} fluxes by 1 magnitude only resulted in $\Delta \chi^2 = 2.9$ and -0.6 compared to the real data.  Even when the {\it HST} fluxes were artificially increased by 3 magnitudes, the model could not be formally rejected at $3 \sigma$ significance.  Notably, the only parameter that changed significantly in these simulations was the reprocessing fraction.  At the current data quality, the other free parameters are determined by the X-ray spectrum. 
Despite this, from an examination of the model residuals at optical wavelengths (Figure \ref{residuals1}) it seems that a larger outer disc radius can better reproduce the observed spectral shape in most of the sources.  Indeed, \cite{gierlinski_etal_2009} note that for the {\sc diskir} model (which neglects the colour-temperature-correction) it is evident from the spectral shape that the irradiated fraction is the main parameter controlling the amount of optical flux, whilst the spectral slope is determined predominantly by the outer disc radius.  An outer disc radius of $10^6~r_{\rm g}$ for a 10--30 $\Msun$ black hole is equal to $\sim$ (1.5--4.5) $\times 10^{12}~{\rm cm}$, this is similar to the typical outer disc radii reported for the ULX sample of \cite{tao_etal_2011}, and is comparable to the $\sim 14~\Msun$ \citep{greiner_etal_2001} mildly super-Eddington \citep{done_etal_2004} Galactic BHB GRS 1915+105, which has an outer disc radius of $\sim 3 \times 10^{12}~{\rm cm}$ and may itself contain an irradiated disc \citep{rahoui_etal_2010}.  In order to accrete via Roche-lobe overflow, such large disc radii would require highly evolved (or massive) companion stars, and they imply relatively long orbital periods for the binary system - e.g., $\sim 10$ days for a $\sim 10~\Msun$ black hole and $\sim 1~\Msun$ companion star, with the orbital period increasing further if the companion star is more massive \citep{tao_etal_2011}.

Next, we consider the reprocessing fractions in the ULX sample.  The best fitting irradiated, colour-temperature-corrected disc models indicate that reprocessing fractions of $\sim 10^{-3}$ are required in NGC 2403 X-1, M81 X-6 and NGC 4190 ULX1, and  $\sim 3 \times 10^{-2}$ in NGC 1313 X-2, although models with a slightly larger black hole and a reprocessing fraction $\sim 10$ times lower cannot be rejected for NGC 1313 X-2.  For the most part, these are a factor of $\sim 10$ lower than the reprocessing fractions claimed for ULXs using the {\sc diskir} model, e.g. $f_{\rm out} = 2.7^{+2.3}_{-1.0} \times 10^{-2}$ in Ho II X-1, \citealt{tao_etal_2012}; $f_{\rm out} \approx (3.0$--$4.6) \times 10^{-2}$ in NGC 5408 X-1, \citealt{grise_etal_2012}.  
However, both Ho II X-1 and NGC 5408 X-1 were classified with soft ultraluminous spectra by \cite{sutton_etal_2013b}, which they suggested was associated with a strong, radiatively-driven wind.  In this case, the increased reprocessing fraction may be in some way triggered by the wind.  Then, the reprocessing fraction of $\sim 3 \times 10^{-2}$ in NGC 1313 X-2 may indicate that this source is actually similar to the more luminous ULXs, and has a strong outflow, which is orientated away from our line-of-sight.  Indeed, \cite{sutton_etal_2013b} cautioned that there was some ambiguity in their spectral classification between the `hard ultraluminous' sources with distinct high energy spectral curvature but little soft excess, and the broad, disc-like ULXs.

In general, the inferred reprocessing fractions in this ULX sample appear to be comparable with those seen in the Galactic BHB XTE J1817$-$330, where the {\sc diskir} model indicates $f_{\rm out} \sim 1 \times 10^{-3}$ \citep{gierlinski_etal_2009}.  Here, to allow for a consistent comparison between our sample and a sub-Eddington Galactic BHB, we extend the analysis of \cite{gierlinski_etal_2009} and fit their spectrum 1 of  XTE J1817$-$330 with the absorbed/reddened {\sc optxirr} model.  The absorption, Eddington ratio, spin, coronal radius and outer disc radius were allowed to vary freely; all of the flux emitted within the coronal radius was approximated as a power-law with the spectral index set to the best fit value from \cite{gierlinski_etal_2009}, $\Gamma = 2.34$; the black hole mass was fixed at 6 $\Msun$, corresponding to the upper limit calculated by \cite{sala_etal_2007}; and, the distance was fixed to 8 kpc, which is consistent with the 1--10 kpc range estimated by \cite{sala_etal_2007}.  We found an absorption column of $\sim 1.5 \times 10^{21}~{\rm cm^{-2}}$, which is greater than the value from the {\sc diskir} fits of \cite{gierlinski_etal_2009}, but is close to the average Galactic column in the direction of the source of $1.57 \times 10^{21}~{\rm cm^{-2}}$ \citep{dickey_and_lockman_1990}.  The other fitted model parameters were an Eddington ratio of $\sim 0.3$, a spin of $\sim 0$, and an outer disc radius of $\sim 1.3 \times 10^5~r_{\rm g}$, which is similar to the value found by \cite{gierlinski_etal_2009} of $10^{4.5}~r_{\rm in}$, assuming an inner disc radius of $6~r_{\rm g}$.  Finally, we estimated an irradiated fraction of $f_{\rm out} \sim 3 \times 10^{-3}$ in XTE J1817$-$330; thus, confirming that it is of the same order of magnitude as the range seen in the ULX sample, even when the colour-temperature-correction is included.

Given the similarities between the reprocessing fractions in these ULXs and the sub-Eddington XTE J1817$-$330, then perhaps some of the ULXs may too be sub-Eddington.  We note that the best fitting {\sc optxirr} models for NGC 2403 X-1 and M81 X-6 indicate that they have Eddington ratios of $\sim$ 0.2--0.3.  But, the cool, optically thick coronae that are seemingly required by the data are unlike any known sub-Eddington state.  Instead, they are indicative of broad, disc-like X-ray spectra.  
The apparent discrepancy between the luminosity and the shape of the X-ray spectrum in these sources may imply that we are over-estimating the masses of the black holes.  If we reject the statistically best fitting models with 30 $\Msun$ black holes in NGC 2403 X-1 and M81 X-6 on physical grounds, then the spectral fits indicate that all four of the ULXs remaining in the sample may be powered by accretion on to $\sim 10~\Msun$ sMBHs, with Eddington ratios of $\sim$ 0.7--2.  
However, we caution that \cite{dotan_and_shaviv_2011} found that the onset of characteristic super-Eddington-like behaviour actually occurs as the Eddington rate is approached.  So, plausibly they may contain MsBH primaries, with a fraction of the flux being lost to advection and mass loss in a wind.  This could result in us under-estimating the global mass accretion rate, which would further exacerbate the problem of the sources appearing to be sub-Eddington.

We suggest that two opposing mechanisms may determine the fraction of the emission reprocessed in the outer disc.  Firstly, ULX accretion discs may be geometrically thick, which would reduce the reprocessing fraction due to self-shielding in the disc \citep{kaaret_and_corbel_2009}.  Secondly, as the Eddington limit is approached and exceeded, accretion discs are expected to drive an outflowing wind \citep{poutanen_etal_2007}.  Material in the wind will be optically thin far from the ULX, and may scatter radiation from the central accretion disc.  Some fraction of this will irradiate the outer disc, thus opposing the effect of self-shielding.  We note that the spectral modelling of \cite{middleton_etal_2014} suggests that the wind will become optically thin at greater than $\sim 10^{3}~r_{\rm g}$, which is consistent with this scenario, for both outer disc radii that we consider here.  Then, the reprocessing fractions here suggest that the effects of these opposing influences must be similar in magnitude in the disc-like ULXs.  Furthermore, this scenario may be compatible with the larger reprocessing fractions reported for some soft ultraluminous ULXs (\citealt{grise_etal_2012}; \citealt{tao_etal_2012}; albeit using the {\sc diskir} model, with no colour-temperature-correction), if a stronger wind results in more of the hard flux from close to the inner disc being scattered.  Indeed, the higher reprocessing fraction in NGC 1313 X-2 may be produced in this way, although this source is likely observed close to face on, so we do not see a soft, wind-dominated X-ray spectrum.  Alternatively, the reprocessing fractions in this sample of ULXs and the thermal dominant state Galactic BHBs may indicate that they share similar reprocessing mechanisms.  In this case, these disc-like ULXs which we suspect are accreting at around the Eddington limit would not have particularly inflated, geometrically thick inner accretion discs. 

There are however a number of caveats to this work.  Firstly, there are inevitably large uncertainties in the reprocessing fractions, as the {\it XMM-Newton} and {\it HST} observations are not simultaneous, and in the cases of M81 X-6 and NGC 4190 ULX1 the {\it HST} observations are themselves spread over time scales of $\sim$ months to a decade.  Clearly, contemporaneous high quality X-ray and optical observations are required to better test the {\sc optxirr} model (and other models that span similar wavelength ranges) in ULXs.  Furthermore, multiple epochs of multi-wavelength data spanning a range of X-ray fluxes would allow us to begin to better constrain some of the degeneracies in e.g. black hole mass inherent in the model.  Also, it should be noted that the spectral model conserves the mass accretion rate in the disc.  This assumption should not be valid in super critical accretion discs, where a significant degree of mass loss is expected to occur.  
Additionally, the model does not include the effects of advection, although this is not expected to be a strong effect at luminosities around the Eddington limit \citep{sadowski_2011}.  Critically, both of these processes would reduce the observed X-ray flux.  So, by not including these effects, we may be underestimating the global mass accretion rate, which is largely determined by the X-ray data.  Thus, we may overestimate the reprocessing fraction required to produce the observed optical flux.  As such, the inclusion of advection and outflows in the model would be expected to reduce the estimated reprocessing fraction, not raise it to the values seen in more luminous ULXs.  
Finally, we have been neglecting the contribution to the optical flux from the companion star.  Given the similar colours of a blue star and emission from the accretion disc, it is difficult to disentangle these two contributions with the available data, although the optical counterparts to some ULXs are too bright to be powered by even the most luminous stars (e.g. \citealt{gladstone_etal_2013}).  Searches for associated X-ray and optical variability would be able to provide a test for reprocessed emission from the outer disc, and such observing campaigns should be possible using currently available observatories.

\section{Conclusions}

We have modelled {\it XMM-Newton} and {\it HST} data from a small sample of ULXs that were identified as having disc-like X-ray spectra indicative of $L_{\rm bol}/L_{\rm Edd} \sim 1$, using the new {\sc optxirr} spectral model, which accounts for: the colour-temperature-correction in the accretion disc; X-ray reprocessing in the outer disc; and a Compton corona covering the inner regions of the disc.  
Although the quality of the available data is such that the X-ray spectra dominate the fit statistics, we are able to make a number of conclusions.  
Contrary to other studies which use the {\sc diskir} model (which neglects the colour-temperature-correction), we find that large reprocessing fractions are not required in several ULXs with disc-like X-ray spectra, rather they are similar in this regard to thermal dominant state Galactic BHBs.  One source -- NGC 1313 X-2 -- may have a higher reprocessing fraction ($\sim 3$ per cent), and this source may be in a wind-dominated super-Eddington accretion state, but with the outflow directed away from our line-of-sight.  Despite the similar reprocessing fractions, the X-ray spectra arguably do not appear to be consistent with these ULXs being very sub-Eddington.  Instead we suggest that the relative effects of two opposing mechanisms may influence the fraction of the bolometric luminosity reprocessed in the outer disc: it can be reduced by self-shielding in the accretion disc; and, material in a nascent super-Eddington disc wind may scatter radiation from above the plane of the disc, such that a fraction of it is reflected on to the outer regions of the accretion disc.  This scenario may be consistent with the larger reprocessing fractions reported for wind-dominated soft ultraluminous ULXs.  Alternatively, the broadened disc ULXs may be similar in this regard to the thermal dominant state BHBs, and accretion rates around the Eddington limit are not sufficiently high to produce particularly geometrically thick accretion discs.

\section*{Acknowledgements}
The authors thank the anonymous referee for their useful comments. They also acknowledge funding from the Science and Technology Facilities Council as part of the consolidated grants ST/K000861/1 and ST/L00075X/1.  This work is based on observations obtained with {\it XMM-Newton}, an ESA science mission with instruments and contributions directly funded by ESA Member States and NASA.  It is also based on observations made with the NASA/ESA {\it Hubble Space Telescope}, obtained at the Space Telescope Science Institute, which is operated by the Association of Universities for Research in Astronomy, Inc., under NASA contract NAS 5-26555.

\bibliography{refs}
\bibliographystyle{mn2e}

\bsp

\label{lastpage}

\end{document}